\documentclass{sf2a-conf2024}
\usepackage{graphicx}
\usepackage{hyperref}
\usepackage[]{natbib}  
\usepackage{epstopdf}

\def\BibTeX{{\rm B\kern-.05em{\sc i\kern-.025em b}\kern-.08em
    T\kern-.1667em\lower.7ex\hbox{E}\kern-.125emX}}
\bibpunct{(}{)}{;}{a}{}{,}  

\usepackage{color}

\newcommand {\el} {\mathrm{el}}

\newcommand{\Lagrange}{Universit{\'e} C{\^o}te d'Azur, Observatoire de la C{\^o}te d'Azur, CNRS, Laboratoire Lagrange, France}
\newcommand{\INPHYNI}{Universit{\'e} C{\^o}te d'Azur, CNRS, Institut de Physique de Nice, France}

\begin{document}

\TitreGlobal{SF2A 2024}


\title{Increasing the sensitivity of stellar intensity interferometry with optical telescopes: first laboratory test of spectral multiplexing}

\runningtitle{Intensity interferometry with optical telescopes}

\author{S. Tolila}\address{\INPHYNI}
\author{G. Labeyrie$^{1}$}
\author{R. Kaiser$^{1}$}
\author{J.-P. Rivet}\address{\Lagrange}
\author{W. Guerin$^{1}$}




\setcounter{page}{237}


\maketitle


\begin{abstract}
We present a preliminary laboratory test of a setup designed to measure Hanbury Brown and Twiss-type intensity correlations from a chaotic light source using five spectral channels simultaneously. After averaging the zero-delay correlation peaks from all channels, we obtain an improvement of the signal-to-noise ratio fairly consistent with theory. The goal is to demonstrate the feasibility and scalability of this technique to improve the sensitivity of stellar intensity interferometry using optical telescopes.  

\end{abstract}



\section{Introduction}

The study of light intensity correlations for astronomical observations was pioneered by Hanbury Brown and Twiss \citep{HBT:1956}. They demonstrated that the amplitude of the ``bunching peak'' $g^{(2)}(r,\tau=0)$ observed when correlating the light detected by two telescopes separated by a projected baseline $r$ is proportional to the squared modulus of the visibility. This technique called stellar intensity interferometry (SII) is relatively insensitive to the phase of the light and thus much easier to implement than amplitude interferometry (SAI), where the lights of the telescopes are directly recombined to observe interference fringes, which requires the control of the path-length difference to better than the optical wavelength. However, the price to pay is the very low sensitivity of SII compared to SAI, which limited so far its use to the observation of very bright stars \citep{HBT:1974}. Nevertheless, the recent progress in the detectors, digital electronics and other photonics technologies has triggered a revival of this technique.

One promising way to implement SII is to use Cherenkov telescope arrays such as CTAO (see, e.g., \citet{Dravins:2013}, and references therein), with very large collectors and a large number of baselines. Recently, successful measurements have been performed with Cherenkov telescopes, namely with the MAGIC \citep{Acciari:2020,MAGIC:2024}, VERITAS \citep{Abeysekara:2020, Acharyya:2024}, and H.E.S.S. arrays \citep{Zmija:2023}.

Our group follows a different approach, based on the use of optical telescopes \citep{Rivet:2018}. There, the quality of the point spread-function (PSF) allows to conveniently couple the light in optical fibers, to use high quantum efficiency detectors with a high timing resolution. Narrow filters can be used allowing for a higher contrast of the bunching peak, to select emission lines, and to remove the contribution of the sky background. However, the sensitivity of the technique has to be greatly increased. This can be achieved by wavelength multiplexing, i.e., by  performing simultaneous independent SII measurements with a large number $N$ of narrow spectral channels. After summarizing our previous achievements, we present here a preliminary laboratory test with a small number of channels ($N = 5$), demonstrating the potential of this scheme.    	
  
\section{Previous achievements}

Our group started its activities on SII in 2017, after using the technique of intensity correlations to study light scattered by hot atomic vapors \citep{Dussaux:2016}. Since then, we have obtained some significant results summarized below.

\subsection{SII with optical telescopes}

Taking advantage of the proximity of the astronomic facility at Calern observatory, we were able to observe first the bunching peak in  stellar light with a single telescope \citep{Guerin:2017}, and then to perform spatial intensity interferometry with two telescopes \citep{Guerin:2018}. These observations were the first intensity interferometry measurements on stars (other than the sun) since the Narrabri Observatory \citep{HBT:1974}, and the first ever using a detection in the photon counting regime. These results triggered a renewed interest in SII with optical telescopes and encouraged other teams to follow this path \citep{Leopold:2024, Walter:2024}.

 \subsection{SII on the H$\alpha$ emission line}

Using narrow filters centered on emission lines, one can isolate fine spectral features and access accurate information on the physics of stars. We performed SII measurements on the H$\alpha$ line at $\lambda = 656.5$\,nm for several stars : P\,Cygni \citep{Rivet:2020,deAlmeida:2022}, Rigel \citep{deAlmeida:2022} and $\gamma$\,Cas \citep{Matthews:2023}. Our results confirmed those obtained earlier by amplitude interferometry and could even be used to slightly refine P\,Cygni's distance.

\subsection{Portability and adaptability of the instrument}

We designed our SII apparatus such that it could be easily transported and adapted to large telescopes in various facilities. To demonstrate this, we undertook several observation campaigns with different instruments : a first one on the 4.1\,m telescope at SOAR in Chile \citep{Guerin:2021}, a second one at Calern Observatory with the MéO 1.5\,m telescope and a portable 1\,m telescope \citep{Matthews:2023}, and two missions at the VLTI (Chile), first using two Auxiliary Telescopes (ATs) on their maintenance station (49\,m baseline) \citep{Matthews:2022}, and more recently three ATs.

\section{First preliminary lab tests towards multichannel measurements}

Intensity interferometry suffers from a low sensitivity compared to traditional (amplitude) interferometry. The signal-to-noise ratio (SNR) of the squared visibility measurement can be written as
\begin{equation}
SNR = \alpha V^2 N_\mathrm{ph}(\lambda) A \sqrt\frac{T_\mathrm{obs}}{4\pi \tau_\el} \,
\end{equation}
where $\alpha$ is an overall photon detection efficiency that includes the detector's quantum efficiency and all losses, $V^2$ is the squared visibility, $N_\mathrm{ph}(\lambda)$ is the incident spectral photon flux, $A$ is the light collection area, $T_\mathrm{obs}$ is the observation time, and $\tau_\el$ the electronic time resolution of the detection (rms width). This expression assumes that the detection is shot-noise limited. To increase the SNR one needs to : increase the size of the telescopes, increase the observation time, and decrease the electronic response time (which determines the width of the bunching peak). Even though the SNR is independent of the optical bandpass, the contrast of the bunching peak is given by the ratio of the light coherence time (determined by the optical bandpass) to the electronic response time. Thus, one needs to use narrow filters (with typical bandpass of the order of 1 nm) to observe a sizeable bunching peak. A natural way to increase the SNR is thus to perform wavelength multiplexing and simultaneously measure intensity correlations in $N$ spectral channels. The SNR is then expected to increase as $\sqrt N$ for a given observation time, or the observation time to decrease as $1/N$ for a given targeted SNR.

\begin{figure}[ht!]
 \centering
\includegraphics[width=0.6\textwidth,clip]{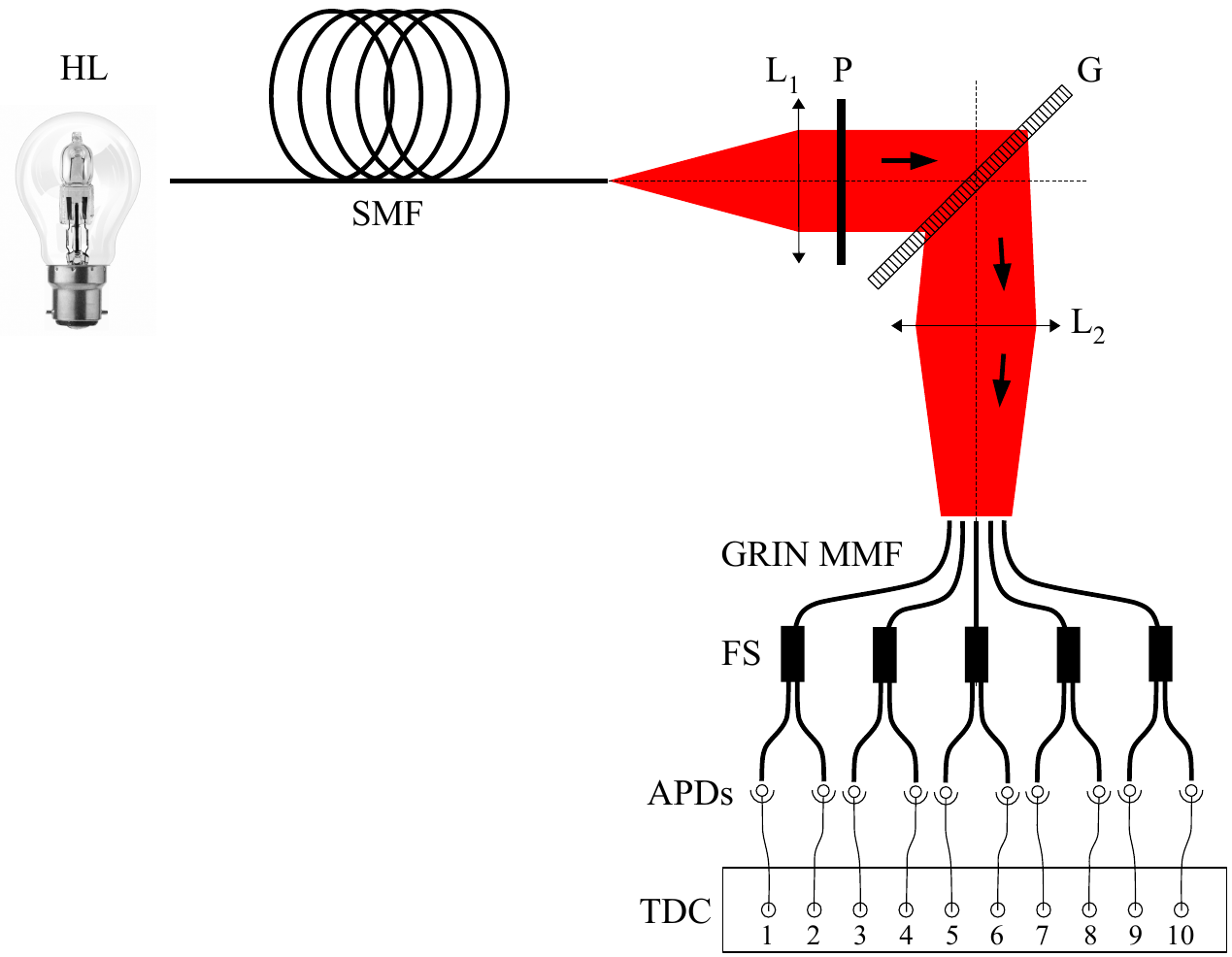}    
  \caption{Laboratory demonstration setup for intensity interferometry with spectral multiplexing.}
  \label{fig.setup}
\end{figure}

We have performed a preliminary laboratory demonstration of intensity correlations with spectral multiplexing, with a modest value of $N = 5$. The architecture of the setup is largely based on the technology used in our previous works \citep{Guerin:2017,Guerin:2018}, e.g. the use of multimode fibers and fiber-coupled avalanche photodiodes (APDs) for detection. A schematic of the setup is shown in Fig.\,\ref{fig.setup}.

We simulate an unresolved star using a halogen lamp (HL) coupled into a single mode fiber (SMF). The light beam delivered by the fiber is collimated using lens L$_1$ (focal length 100 mm). The collimated beam diameter is $\sim 1$\,cm. The polarization of the light is selected using polarizer P. To achieve multiple spectral channels, we disperse the light using a transmission grating (Wasatch Photonics) with a high diffraction efficiency of $95\%$ and an angular dispersion $D = 3.06$ mrad/nm around $\lambda = 780$\,nm. The angular (far-field) distribution of the light diffracted by the grating is formed in the focal plane of lens L$_2$ (focal length $f = 100$ mm). In this focal plane, we place a bundle of ten graded-index multimode fibers (GRIN MMF) with a core diameter $\Phi = 100\,\mu$m and a numerical aperture of 0.29 (Berkshire Photonics). This numerical aperture is larger than that of the incoming beam ($\approx 0.1$), ensuring optimal coupling. The fiber's input facets are aligned on a line parallel to the direction of diffraction by the grating. A rough estimate of the effective bandwidth of each spectral channel is $\Delta \lambda = \Phi/f\,D = 0.33$\,nm.
Because of a limited available number of ten detectors, we use only five of the ten bundled fibers. Each of these is connected to a 50/50 fibered splitter (FS) whose outputs are sent on APDs (Excelitas Technologies), allowing us to obtain temporal intensity correlation signals from each channel. The temporal jitter of the ADPs leads to an electronic response time $\tau_\el \simeq 900$\,ps (FWHM). All the outputs of the ten APDs are sent to a time-to-digital convertor (TDC, Swabian Instruments) connected to a computer. We compute and store the five correlation signals in real time.

\begin{figure}[ht!]
 \centering
 \includegraphics[width=0.7\textwidth,clip]{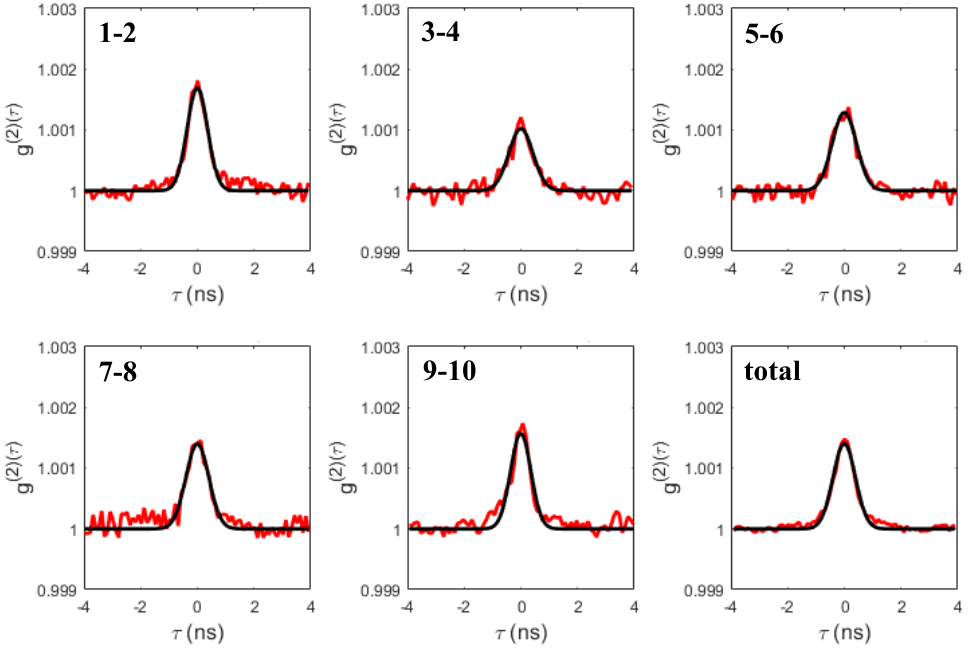}    
  \caption{Temporal intensity correlations measurement with five spectral channels.}
  \label{fig.results}
\end{figure}

The result of this experiment is shown in Fig.\,\ref{fig.results}. The measurement was performed with approximately $2\times10^6$ counts per second on each detector. The signals were integrated for 69 hours. We show in Fig.\,\ref{fig.results} the five correlation peaks, corresponding to each spectral channels, and the resulting average (``total''). As expected, the SNR is improved by the averaging. We extract from the Gaussian fits the SNR in the different channels : 32 (1-2), 22 (3-4), 30 (5-6), 21 (7-8), and 25 (9-10). The SNR for the averaged signal is 51, slightly below the expected value 59 = $\sqrt {32^2 + 22^2 + 30^2 + 21^2 + 25^2}$.

We also observe that bunching peaks in different channels have different contrasts and widths. This is due to the convolution of the bunching peaks by the temporal responses of the detectors, which are different for each channel. However the area under the correlation peak, which gives the coherence time $\tau_\mathrm{c}$, is not affected by this convolution. There are all consistent with one another ($\tau_\mathrm{c}\simeq 1.45$\,ps), except for the second pair, which is  lower ($\tau_\mathrm{c}\simeq 1$\,ps). This might indicate some variations among the fibers within the bundle, inducing a different effective spectral filtering. Even more striking, the measured coherence time is quite lower than expected. 
Indeed, for a square spectrum of width $\Delta \lambda =0.33$\,nm, we get $\tau_c = 6\,$ps.
We believe this is due to the specificity of light coupling into GRIN fibers: the coupling efficiency is space dependent, creating a complicated spectral profile, which changes the coherence time. 
This effect remains to be studied in more detail, for instance by directly measuring the spectral profiles associated with each channel using a high-resolution spectrometer, as we have previously done for our narrow filters \citep{Matthews:2022}.

\section{Conclusions}

We presented in this paper a preliminary lab test on simultaneous temporal intensity correlations with several spectral channels. We demonstrated the efficiency of the method and the subsequent improvement of the SNR. Some limitations and open questions have also been identified, which should be the subject of further investigations, before extending this method to large $N$ numbers up to several hundreds and more. The advent of large arrays of e.g. Single Photon Avalanche Detectors (SPAD) allows to foresee such a development. The advances in photonic devices such as photonic lanterns and arrayed waveguide gratings could potentially lead to very compact and stable devices \citep{Lai:2018}. In the end, a stringent requirement for a future apparatus is that the overall photon detection efficiency (the $\alpha$ term in Eq.\,1) remains as high as possible, such that the gain of SNR due to increasing $N$ is fully achieved.   

\begin{acknowledgements}
We acknowledge funding from the ANR (project I2C, ANR-20-CE31-0003). 
\end{acknowledgements}



%
\end{document}